\documentclass[a4paper,11pt]{article}
\usepackage[ams]{}
\usepackage{amsmath}
\usepackage{amsthm}
\usepackage{hyperref}
\usepackage{amssymb}
\usepackage{graphicx}
\usepackage[all,arc]{xy}
\usepackage[theorem]{}

\newcommand{\norm}[1]{\lVert#1\rVert}

\theoremstyle{plain}			% use "default" font

\newtheorem{thm}{Theorem}[section]

\theoremstyle{definition}		% use "definition-style" font for the rest

\begin{document}
\title{Extremal Black Holes as Qudits}
\author{Michael Rios\footnote{email: mrios4@calstatela.edu}\\\\\emph{California State University, Los Angeles}\\\emph{Physics Graduate Program}
\\\emph{5151 State University Drive}\\\emph{Los Angeles, CA 90032, USA}  } \date{\today}\maketitle
\begin{abstract}
We extend the black hole/qudit correspondence by identifying five and six-dimensional 1/2-BPS black string and hole charge vectors in $\mathcal{N}=8$ and $\mathcal{N}=2$ magic supergravities with qubits and qutrits over composition algebras.  In $D=6$, this is accomplished via Hopf fibrations, which map qubits over composition algebras to rank one elements of Jordan algebras of degree two.  An analogous procedure maps qutrits over composition algebras to $D=5$ charge vectors, which are rank one elements of Jordan algebras of degree three.  In both cases, the U-duality groups are interpreted as qudit SLOCC transformation groups. We provide explicit gates for such transformations and study their applications in toroidally compactified M-theory.
\\\\
$Keywords:$ Black Holes, Qudits, U-duality.
\end{abstract}

\newpage
\tableofcontents
\section{Introduction}
\indent In recent papers \cite{1}-\cite{9}\cite{55}-\cite{57}, Ferrara, Lev\'{a}y and Duff et al. established a correspondence between extremal black holes and qubit and qutrit entanglement.  Such a correspondence stems from Duff's observation that the entropy of an extremal black hole in the four-dimensional $\mathcal{N}=2$ STU model and the measure of three-qubit entanglement are both given by Cayley's hyperdeterminant \cite{8}.  Later work was motivated by studies of BPS black holes in a class of $\mathcal{N}=2$ supergravities whose scalar fields lie on a symmetric space \cite{10}-\cite{14}, called magic supergravities \cite{15}-\cite{18}.  It was shown that extremal black holes in $D=3,4,5,6$ magic supergravities can be described by Jordan algebras and their corresponding Freudenthal triple systems, with the Bekenstein-Hawking entropy of the black holes given by algebraic invariants \cite{4,10,13,14,19,20}.  In $D=4$, the quartic invariant was shown to be proportional to Cayley's hyperdeterminant \cite{2,3,4} and used to describe three-qubit entanglement, yielding a novel derivation of the three-qubit entanglement classes \cite{5}.  Subsequent work led to the description of STU black holes in terms of four qubits \cite{57} and a rigorous classification of four qubit entanglement based on $D=3$ black hole U-duality orbits \cite{55,56}. 

\indent In \cite{4}, Borsten et al. proposed the identification of elements of Jordan algebras of degree two and three with $2 \times 2$ and $3 \times 3$ reduced density matrices for two qubits and qutrits, in quantum mechanics defined over arbitrary division algebras.  Here, we also make use of the Jordan formulation of quantum mechanics, but give a different interpretation for elements of Jordan algebras over composition algebras, which include division algebras as well as their split forms.  Given that 1/2-BPS black hole charge vectors in $D=5,6$, $\mathcal{N}=8$ and $\mathcal{N}=2$ magic supergravities are rank one elements of Jordan algebras of degree two and three \cite{13,52}, it is possible to identify 1/2-BPS black hole charge vectors with pure qudits (qubits and qutrits) over composition algebras.  The unitary and SLOCC groups of such qudits correspond to the automorphism and reduced structure groups of the corresponding Jordan algebras, respectively.  As higher rank Jordan elements can always be expressed as orthogonal sums of rank one elements via spectral decomposition \cite{26,29}, we identify 1/4 and 1/8-BPS black hole charge vectors with observables on orthogonal qudit states. 
  
Geometrically, 1/2-BPS black hole charge vectors are mapped to points in projective space, while 1/4 and 1/8-BPS charge vectors are mapped to projective lines and planes.  The corresponding U-duality groups are isomorphic to the collineation groups of the projective spaces \cite{24}.  In the quantum information context, U-duality transformations are realized by acting on Jordan algebra matrices with elements of the reduced structure group \cite{45}, where such elements are interpreted as qudit SLOCC gates.

\section{$\mathcal{N}=8$ and $\mathcal{N}=2$ Magic Supergravities}
\indent We review $\mathcal{N}=8$ and $\mathcal{N}=2$ magic supergravities in $D=5,6$ and give their corresponding Jordan algebras, symmetry groups and entropy expressions for black string and hole solutions.  The relevant Jordan algebras are of degree two and three and taken over finite dimensional composition algebras.  We will first review composition algebras and Jordan algebras, particularly those relevant to black hole solutions in supergravity.  
\subsection{Composition Algebras}
Let $V$ be a finite dimensional vector space over a field $\mathbb{F}=\mathbb{R},\mathbb{C}$.  An $algebra$ $structure$ on $V$ is a bilinear map
\begin{eqnarray}
V\times V \rightarrow V \\
(x,y)\mapsto x\bullet y.
\nonumber
\end{eqnarray}
A $composition$ $algebra$ is an algebra $\mathbb{A}=(V,\bullet)$, admitting an identity element, with a non-degenerate quadratic form $\eta$ satisfying
\begin{eqnarray}
\forall x,y\in\mathbb{A}\quad \eta(x\bullet y)=\eta(x)\eta(y).
\end{eqnarray}
If $\exists x\in\mathbb{A}$ such that $x\neq 0$ and $\eta(x)=0$, $\eta$ is said to be $isotropic$ and gives rise to a $split$ $composition$ $algebra$.  When $\forall x\in\mathbb{A}$, $x\neq 0$, $\eta(x)\neq 0$, $\eta$ is $anisotropic$ and yields a $composition$ $division$ $algebra$.
\begin{thm}
A finite dimensional vector space $V$ over $\mathbb{F}=\mathbb{R},\mathbb{C}$ can be endowed with a composition algebra structure if and only if $\textrm{dim}_{\mathbb{F}}(V)=1,2,4,8$.  If $\mathbb{F}=\mathbb{C}$, then for a given dimension all composition algebras are isomorphic.  For $\mathbb{F}=\mathbb{R}$ and $\textrm{dim}_{\mathbb{F}}(V)=8$ there are only two non-isomorphic composition algebras: the octonions $\mathbb{O}$ for which $\eta$ is anisotropic and the split-octonions $\mathbb{O}_s$ for which $\eta$ is isotropic and of signature $(4,4)$.  Moreover for all composition algebras, the quadratic form $\eta$ is uniquely defined by the algebra structure.
\end{thm}

The composition algebras immediately applicable to $\mathcal{N}=8$ and $\mathcal{N}=2$ magic supergravities include those over $\mathbb{R}$ with anisotropic quadratic form, the $\textrm{dim}_{\mathbb{R}}(V)=1,2,4,8$ division algebras $\mathbb{R},\mathbb{C},\mathbb{H},\mathbb{O}$, and the eight-dimensional split octonions $\mathbb{O}_s$ with isotropic quadratic form.  It is worth noting that even though $\mathbb{O}$ and $\mathbb{O}_s$ are non-isomorphic, they are both real subalgebras of the unique $\textrm{dim}_{\mathbb{C}}(V)=8$ dimensional bioctonion composition algebra $\mathbb{O}_{\mathbb{C}}$ \cite{19,54}.

\subsection{Jordan Algebras}

\indent A Jordan algebra over a field $\mathbb{F}=\mathbb{R},\mathbb{C}$ is a vector space over $\mathbb{F}$ equipped with a bilinear form (Jordan product) $(X,Y)\rightarrow X\circ Y$ satisfying $\forall X,Y\in\mathcal{A}$:
\begin{displaymath}
X\circ Y = Y\circ X
\end{displaymath}
\begin{displaymath}
X\circ (Y\circ X^2)=(X\circ Y)\circ X^2.
\end{displaymath}
Given an associative algebra $\mathcal{A}$, we can define the Jordan product $\circ$ on $\mathcal{A}$ using the associative product in $\mathcal{A}$, given by $X\circ Y\equiv \frac{1}{2}(XY+YX)$.  Under the Jordan product, ($\mathcal{A},\circ$) becomes a \emph{special} Jordan algebra $\mathcal{A}^{+}$.  Any Jordan algebra that is simple and not special is called an \emph{exceptional} Jordan algebra or Albert Algebra.  The exceptional Jordan algebras are 27-dimensional and over $\mathbb{R}$ are isomorphic to either $J^{\mathbb{O}}_3$ or $J^{\mathbb{O}_s}_3$ \cite{26}. 

\indent An idempotent (or projector) in a Jordan algebra is a non-zero element such that $P^2=P$.  Given two idempotents $P_1,P_2$, they are orthogonal if $P_1P_2=0$. An idempotent is called \emph{primitive} if it is not the sum of two orthogonal idempotents.  In finite dimensions, any idempotent can be decomposed into an orthogonal sum of primitive idempotents.  The number of primitive idempotents the identity decomposes into is called the \emph{capacity}, and determines the \emph{degree} of the Jordan algebra \cite{25}.

The Jordan algebras relevant to $D=5,6$, $\mathcal{N}=8$ and $\mathcal{N}=2$ magic supergravities are those of degree two and three, consisting of $2 \times 2$ and $3\times 3$ Hermitian matrices with entries in a composition algebra.  The automorphism and determinant preserving groups for these algebras will be denoted by $\textrm{Aut}(J^{\mathbb{A}}_n)$ and $\textrm{Str}_0(J^{\mathbb{A}}_n)$, respectively.

\subsection{$D=6$ Black Strings}
\indent The $D=6$, $\mathcal{N}=8$ and $\mathcal{N}=2$ magic supergravities arise as uplifts of $D=5$, $\mathcal{N}=8$ and $\mathcal{N}=2$ magic supergravities with $n_V=27$, $n_V=15$, $n_V=9$ and $n_V=6$ vector fields \cite{21}.  They exhibit $\textrm{Spin}(5,5)$, Spin$(9,1)$, Spin$(5,1)$, Spin$(3,1)$ and Spin$(2,1)$ U-duality symmetry since the $D=6$ vector multiplets in the Coulomb phase (after Higgsing) transform as spinors of dimension 16, 8, 4 and 2, respectively \cite{21}.  One can associate a black string solution with charges $q_I$ ($I=1,...,n_V$ for $n_V=10,6,4,3$) an element
\begin{equation}
J=\sum_{I=1}^{n}q^Ie_I=\left(\begin{array}{cc} r_1 & A \\ \overline{A}& r_2  \end{array}\right)\quad r_i\in\mathbb{R}, A\in\mathbb{A}
\end{equation}
of a Jordan algebra $J^{\mathbb{A}}_2$ of degree two, where the $e_I$ form the $n_V$-dimensional basis of the Jordan algebra over a composition algebra $\mathbb{A}$.  Elements of $J^{\mathbb{A}}_2$ transform as the (dim $\mathbb{A}+2$) representation of $SL(2,\mathbb{A})$, the $\mathbf{10}$, $\mathbf{10}$, $\mathbf{6}$, $\mathbf{4}$, $\mathbf{3}$ of $SO(5,5)$, $SO(9,1)$, $SO(5,1)$, $SO(3,1)$ and $SO(2,1)$, respectively \cite{4}.  This can be seen by noting the $(p,q)$ spacetime inner product can be expressed as
\begin{equation}
J\cdot K=\frac{1}{2}(\textrm{tr}(J)\textrm{tr}(K)-\textrm{tr}(J\circ K))
\end{equation}  
and by setting $J=K$, $r_1=x+y$ and $r_2=x-y$ we recover the $(p,q)$ spacetime norm squared
\begin{equation}
J\cdot J=\textrm{det}(J)=x^2-y^2+A\overline{A}.
\end{equation}
As $SL(2,\mathbb{A})=\textrm{Str}_0(J_2^{\mathbb{A}})$ consists of determinant preserving transformations over $J^{\mathbb{A}}_2$,  such transformations necessarily preserve the $(p,q)$ spacetime norm.  Moreover, since $SL(2,\mathbb{O}_s)=\textrm{Spin}(5,5)$, $SL(2,\mathbb{O})=\textrm{Spin}(9,1)$, $SL(2,\mathbb{H})=\textrm{Spin}(5,1)$, $SL(2,\mathbb{C})=\textrm{Spin}(3,1)$ and $SL(2,\mathbb{R})=\textrm{Spin}(2,1)$, the original claim for the $SO(p,q)$ groups follows.

Over any composition algebra, the black string entropy is given by \cite{4}
\begin{equation}
S=\pi\sqrt{|I_2(J)|}
\end{equation}
where
\begin{equation}
I_2(J)=\textrm{det}(J)=r_1r_2-A\overline{A}.
\end{equation}
The U-duality orbits are distinguished by rank \cite{52} via
\begin{equation}\begin{array}{rcl}
\textrm{Rank}\thinspace J=2\quad \textrm{iff}\quad I_2(J)\neq 0\quad\quad\quad\qquad\hspace{1pt} S\neq 0,\thinspace \textrm{1/4-BPS}\hfill\\
\textrm{Rank}\thinspace J=1\quad \textrm{iff}\quad I_2(J)=0,\thinspace J\neq 0\qquad S=0,\thinspace\textrm{1/2-BPS}\hfill.
\end{array}
\end{equation}
Being that 1/2-BPS black hole charge vectors satisfy $I_2(J)=J\cdot J=0$, they transform as lightlike vectors in $(p,q)$ spacetime. 

\subsection{$D=5$ Black Holes}
\indent In $D=5$, the $\mathcal{N}=8$ and $\mathcal{N}=2$ supergravities are coupled to 6, 9, 15 and 27 vector fields with U-duality symmetry groups $SL(3,\mathbb{R})$, $SL(3,\mathbb{C})$, $SU^\ast(6)$, $E_{6(-26)}$ and $E_{6(6)}$, respectively \cite{4,13,14}.  The orbits of BPS black hole solutions were classified \cite{13,22} by studying the underlying Jordan algebras of degree three under the actions of their reduced structure groups, $\textrm{Str}_0(J_3^{\mathbb{A}})$, which correspond to the U-duality groups of the $\mathcal{N}=8$ and $\mathcal{N}=2$ supergravities.  This is seen by associating a given black hole solution with charges $q_I$ ($I=1,...,n_V)$ an element
\begin{equation}
J=\sum_{I=1}^{n}q^Ie_I=\left(\begin{array}{ccc} r_1 & A_1 & \overline{A}_2 \\ \overline{A}_1 & r_2 & A_3 \\ A_2 & \overline{A}_3 & r_3  \end{array}\right)\quad r_i\in\mathbb{R}, A_i\in\mathbb{A}
\end{equation}
of a Jordan algebra of degree three $J_3^{\mathbb{A}}$ over a composition algebra, where $e_I$ form a basis for the $n_V$-dimensional Jordan algebra.  This establishes a correspondence between Jordan algebras of degree three and the charge spaces of the extremal black hole solutions \cite{13}.  In all cases, the entropy of a black hole solution can be written \cite{4,13,23} in the form
\begin{equation}
S=\pi\sqrt{|I_3(J)|}
\end{equation}
where $I_3$ is the cubic invariant given by 
\begin{equation}
I_3=\textrm{det}(J)
\end{equation}
and 
\begin{equation}
\textrm{det}(J)=r_1r_2r_3-r_1||A_1||^2-r_2||A_2||^2-r_3||A_3||^3+2\textrm{Re}(A_1 A_2 A_3).
\end{equation}
The U-duality orbits are distinguished by rank via
\begin{equation}\begin{array}{rcl}
\textrm{Rank}\thinspace J=3\quad \textrm{iff}\quad I_3(J)\neq 0\hspace{3pt}\qquad\quad\qquad S\neq 0,\thinspace \textrm{1/8-BPS}\hfill\\
\textrm{Rank}\thinspace J=2\quad \textrm{iff}\quad I_3(J)=0, J^{\natural} \neq 0\qquad S=0,\thinspace \textrm{1/4-BPS}\hfill\\
\textrm{Rank}\thinspace J=1\quad \textrm{iff}\quad J^{\natural} = 0,\thinspace\thinspace J\neq 0\thinspace\thinspace\quad\qquad S=0,\thinspace\textrm{1/2-BPS}\hfill
\end{array}
\end{equation}
where the quadratic adjoint map is given by
\begin{equation}
J^{\natural}=J\times J = J^2-\textrm{tr}(J)J+\frac{1}{2}(\textrm{tr}(J)^2-\textrm{tr}(J^2))I.
\end{equation}
Note the quadratic adjoint map is a generalization of the $I_2$ invariant, whose vanishing defined 1/2-BPS black hole solutions in $D=6$.  It will be seen that the vanishing of the quadratic adjoint map or the $I_2$ invariant are defining conditions for 1/2-BPS black hole charge vectors to be interpreted as points in projective space.

\section{Qubits, Qutrits and Black Holes}
In this section, qubits and qutrits over composition algebras will be defined and identified with 1/2-BPS black string and hole charge vectors in $D=5,6$, $\mathcal{N}=8$ and $\mathcal{N}=2$ magic supergravities. 
\subsection{Qubits and $D=6$ Black Strings}
A quantum bit or $qubit$ is a state of a quantum system with a two-dimensional representation space $\mathcal{H}^2$.  The basis states of this space are denoted as $|0\rangle$ and $|1\rangle$ and a general state can be written in this basis as:
\begin{equation}
|\Psi\rangle=a_0|0\rangle+a_1|1\rangle\quad\quad a_i\in\mathcal{H}^2.
\end{equation}  
$|\Psi\rangle$ is usually taken to be an element of the Hilbert space $\mathbb{C}^2$, where
\begin{equation}
|0\rangle=\left(\begin{array}{c} 1 \\ 0 \end{array}\right)\quad|1\rangle=\left(\begin{array}{c} 0 \\ 1 \end{array}\right)
\end{equation} 
represent the qubit computational basis.  A $pure$ qubit satisfies the normalization condition $||a_0||^2+||a_1||^2=1$, where $||a_i||^2$ is the probability of measuring the qubit in state $|i\rangle$.  Two pure states $|\psi\rangle$ and $|\phi\rangle$ can be obtained with certainty from each other by means of local operations assisted with classical communication (LOCC) if and only if they are related by local unitaries LU \cite{32,33}.  If we merely require that two states $|\psi\rangle$ and $|\phi\rangle$ be obtained from each other with a non-vanishing probability of success, the conversion of the states is performed through stochastic local operations and classical communication (SLOCC) \cite{32,33}.  For a single qubit, the LOCC and SLOCC equivalence groups are equivalent to $SU(2)$ and $SL(2,\mathbb{C})$ \cite{5,35}.\\
\indent Given a pure qubit, one can assign a rank one density matrix
\begin{equation}
P=|\Psi\rangle\langle \Psi|=\left(\begin{array}{cc} ||a_0||^2 & a_0\overline{a}_1 \\ a_1\overline{a}_0 & ||a_1||^2\end{array}\right).
\end{equation}
satisfying
\begin{equation}\begin{array}{ccl}
P^2=P\\
\textrm{tr}(P)=1.
\end{array}
\end{equation}
As $P$ is a rank one $2\times 2$ Hermitian idempotent, we have a map from the Hilbert space $\mathbb{C}^2$ to the space of rank one elements of the Jordan algebra $J^{\mathbb{C}}_2$.  Geometrically, $P$ is mapped to a point of the projective line $\mathbb{CP}^1$ \cite{24}.  In fact, the mapping of a pure qubit to $\mathbb{CP}^1$ gives rise to the Hopf fibration $S^1\hookrightarrow S^3\rightarrow S^2$.  In general, given a pure qubit over a composition divison algebra $\mathbb{A}=\mathbb{R},\mathbb{C},\mathbb{H},\mathbb{O}$, the mapping $\mathbb{A}^2\rightarrow \mathbb{AP}^1$ gives rise to all four Hopf fibrations \cite{24}
\begin{align}
\mathbb{R}^2\rightarrow \mathbb{RP}^1:\qquad\thinspace\thinspace\thinspace\thinspace\thinspace S^0\hookrightarrow S^1\rightarrow S^1\nonumber\\
\mathbb{C}^2\rightarrow \mathbb{CP}^1:\qquad\thinspace\thinspace\thinspace\thinspace\thinspace S^1\hookrightarrow S^3\rightarrow S^2\nonumber \\
\mathbb{H}^2\rightarrow \mathbb{HP}^1:\qquad\thinspace\thinspace\thinspace\thinspace S^3\hookrightarrow S^7\rightarrow S^4 \nonumber \\
\mathbb{O}^2\rightarrow \mathbb{OP}^1:\qquad S^7\hookrightarrow S^{15}\rightarrow S^8. 
\end{align}
For split composition algebras $\mathbb{A}=\mathbb{C}_s,\mathbb{H}_s,\mathbb{O}_s$, there exist analagous mappings \cite{58}, providing non-compact Hopf fibrations which represent maps between hyperboloids in different dimensions with hyperboloid fibers
\begin{align}
\mathbb{C}_s^2\rightarrow \mathbb{CP}_s^1:\qquad\thinspace\thinspace\thinspace H^{1,0}\hookrightarrow H^{2,1}\rightarrow H^{1,1}\nonumber \\
\mathbb{H}_s^2\rightarrow \mathbb{HP}_s^1:\qquad\thinspace\thinspace H^{2,1}\hookrightarrow H^{4,3}\rightarrow H^{2,2} \nonumber \\
\mathbb{O}_s^2\rightarrow \mathbb{OP}_s^1:\qquad H^{4,3}\hookrightarrow H^{8,7}\rightarrow H^{4,4}. 
\end{align}
The non-compact Hopf fibrations thus describe the geometry of qubit state spaces over split composition algebras.  As $SU(2)=\textrm{Aut}(J^{\mathbb{C}}_2)$, it is natural to identify LOCC groups for pure qubits over composition algebras with the automorphism group of the corresponding Jordan algebra of degree two $\textrm{Aut}(J^{\mathbb{A}}_2)$, yielding $SO(2)$, $SU(2)$, $Usp(4)$ and $SO(9)$ for composition division algebras $\mathbb{R}$, $\mathbb{C}$, $\mathbb{H}$, $\mathbb{O}$ \cite{24}, and $SO(2,1)$, $SO(3,2)$ and $SO(5,4)$ for split composition algebras $\mathbb{C}_s$, $\mathbb{H}_s$ and $\mathbb{O}_s$, respectively \cite{58}.  The relevant SLOCC groups are then given by the reduced structure groups $SL(2,\mathbb{A})=\textrm{Str}_0(J_2^{\mathbb{A}})$, giving $SL(2,\mathbb{R})=\textrm{Spin}(2,1)$, $SL(2,\mathbb{C})=\textrm{Spin}(3,1)$, $SL(2,\mathbb{H})=\textrm{Spin}(5,1)$ and $SL(2,\mathbb{O})=\textrm{Spin}(9,1)$ for composition division algebras \cite{24}, and $SL(2,\mathbb{C}_s)=\textrm{Spin}(2,2)$, $SL(2,\mathbb{H}_s)=\textrm{Spin}(3,3)$, $SL(2,\mathbb{O}_s)=\textrm{Spin}(5,5)$ for split composition algebras.

With a direct correspondence between qubit state spaces and spaces of rank one elements of Jordan algebras of degree two, it is seen that six-dimensional 1/2-BPS black hole charge vectors transform as qubits over composition algebras.  This allows one to identify 1/2-BPS black holes with qubits and also gives a quantum computational interpretation for $D=6$, $\mathcal{N}=8$ and $\mathcal{N}=2$ magic supergravitiy U-duality groups. Specifically, U-duality groups consist of those qubit transformations which allow the conversion between states with a non-vanishing probability of success.  This is sensible since, in general, special linear transformations are not isometries of qubit state spaces and not expected to preserve the normalization of pure qubits.  However, such transformations do preserve the determinant, hence rank, acting as single qubit SLOCC gates which map between not necessarily normalized states in projective space.

Single qubit gates are typically given by unitary transformations \cite{36,37}, but for reversible computation any $2\times 2$ invertible linear transformation will suffice \cite{38}.  Important unitary single qubit gates include the Hadamard gate
\begin{equation}
H=\frac{1}{\sqrt{2}}\left(\begin{array}{cc}  1 & 1  \\ 1 & -1 \end{array}\right),
\end{equation}
the Pauli-X,Y,Z gates (where Pauli-X acts as a NOT gate)
\begin{equation}
X=\left(\begin{array}{cc}  0 & 1  \\ 1 & 0 \end{array}\right)\quad Y=\left(\begin{array}{cc}  0 & -i  \\ i & 0 \end{array}\right)\quad Z=\left(\begin{array}{cc}  1 & 0  \\ 0 & -1 \end{array}\right),
\end{equation}
and the phase shift gates
\begin{equation}
R_{\theta}=\left(\begin{array}{cc}  1 & 0  \\ 0 & e^{i\theta} \end{array}\right).
\end{equation}
Some example non-unitary gates \cite{38} include
\begin{equation}
N_1=\left(\begin{array}{cc}  1 & 0  \\ 0 & r_1 \end{array}\right)\quad N_2=\left(\begin{array}{cc}  r_2 & 0  \\ 0 & 1 \end{array}\right)\quad r_i\in\mathbb{R}, r_i\neq 0.
\end{equation}
Over any composition algebra $\mathbb{A}$, the Pauli gates are generalized to the set
\begin{equation}
X=\left(\begin{array}{cc}  0 & 1  \\ 1 & 0 \end{array}\right)\quad Y_i=\left(\begin{array}{cc}  0 & \overline{l}_i  \\ l_i & 0 \end{array}\right)\quad Z=\left(\begin{array}{cc}  1 & 0  \\ 0 & -1 \end{array}\right)\quad i=1,...,\textrm{dim}(\mathbb{A})-1\nonumber
\end{equation}
where $l_i^2=(\textrm{e}_i)^2=-1$ or $l_{i,s}^2=(\textrm{ie}_i)^2=1$ and the phase shift gates become
\begin{equation}
R_{i,\theta}=\left(\begin{array}{cc}  1 & 0  \\ 0 & e^{l_i\theta} \end{array}\right).
\end{equation}
It can be verified that the generalized Pauli gates $X,Y_i$ and $Z$ are generators for $\textrm{Aut}(J_2^{\mathbb{A}})$, while $H,R_{i,\theta}\in\textrm{Aut}(J_2^{\mathbb{A}})$ and $N_k\in\textrm{Str}(J_2^{\mathbb{A}})$.  Using elements of $\textrm{Aut}(J_2^{\mathbb{A}})$ and $\textrm{Str}(J_2^{\mathbb{A}})$ with entries in a complex subalgebra \cite{45}, single qubit gates act on general elements $J\in J_2^{\mathbb{A}}$ via conjugation
\begin{equation}
J'=SJS^{\dagger}.
\end{equation}
Gates in $\textrm{Aut}(J_2^{\mathbb{A}})$ preserve the Frobenius norm $\norm{J}^2=\textrm{tr}(J^2)$ while the $\textrm{Str}(J_2^{\mathbb{A}})$ gates preserve the determinant $\textrm{det}(J)$ up to a real constant.  When the determinant is preserved exactly, one recovers SLOCC gates in $\textrm{Str}_0(J_2^{\mathbb{A}})$.\newline
\indent Given a black string in $D=6$, $\mathcal{N}=8$ or $\mathcal{N}=2$ magic supergravity, its charge vector is an element $J\in J_2^{\mathbb{A}}$ \cite{4} with spectral decomposition
\begin{equation}
J=\lambda_1P_1+\lambda_2P_2\quad\quad \lambda_i\in\mathbb{R},
\end{equation} 
where $P_i$ are orthonormal rank one idempotents of $J_2^{\mathbb{A}}$.  Identifying $P_1,P_2$ with pure qubit density matrices allows one to interpret a 1/4-BPS black string charge vector as a Boolean observable on two orthogonal qubit states.  The eigenvalues $\lambda_i$ are those values the observable takes on two coordinate charts of the projective line $\mathbb{AP}^1$ \cite{51}.  The group of qubit SLOCC transformations $\textrm{Str}_0(J_2^{\mathbb{A}})$ transforms the observable while preserving the black string entropy $S=\pi\sqrt{|\textrm{det}(J)|}=\pi\sqrt{|\lambda_1\lambda_2|}$.  This is consistent with the interpretation of $D=6$ black strings as bound states or intersections of basic constituent $p$-branes of charge $Q_1,Q_2$, for which the area law for the entropy yields $S=\pi\sqrt{|Q_1Q_2|}$ \cite{50}.  These configurations preserve 1/4 of the supersymmetry and in $\mathcal{N}=8$ supergravity can arise from two orthogonally intersecting M-branes in $D=11$ \cite{43,44,46,47,48}.
\subsection{Qutrits and $D=5$ Black Holes}
\indent A quantum trit or $qutrit$ is a state of a quantum system with a three-dimensional representation space.  The basis states of this space are denoted as $|0\rangle$, $|1\rangle$ and $|2\rangle$ and a general state can be written in this basis as:
\begin{equation}
|\Psi\rangle=a_0|0\rangle+a_1|1\rangle+a_2|2\rangle\quad\quad a_i\in\mathcal{H}^3.
\end{equation}  
$|\Psi\rangle$ is usually taken to be an element of $\mathbb{C}^3$, where
\begin{equation}
|0\rangle=\left(\begin{array}{c} 1 \\ 0 \\ 0 \end{array}\right)\quad |1\rangle=\left(\begin{array}{c} 0 \\ 1 \\ 0 \end{array}\right)\quad |2\rangle=\left(\begin{array}{c} 0 \\ 0 \\ 1 \end{array}\right)
\end{equation} 
represent the qutrit computational basis.  A $pure$ qutrit satisfies the normalization condition $||a_0||^2+||a_1||^2+||a_2||^2=1$, where $||a_i||^2$ is the probability of measuring the qutrit in state $|i\rangle$. For a single qutrit, the LOCC and SLOCC equivalence groups are $SU(3)$ and $SL(3,\mathbb{C})$, respectively \cite{3,4,34,39}.\\
\indent Given a pure qutrit, we can construct its rank one density matrix
\begin{equation}
P=|\Psi\rangle\langle \Psi|=\left(\begin{array}{ccc} ||a_0||^2 & a_0\overline{a}_1 & a_0\overline{a}_2 \\ a_1\overline{a}_0 & ||a_1||^2 & a_1\overline{a}_2 \\ a_2\overline{a}_0 & a_2\overline{a}_1 & ||a_2||^2\end{array}\right).
\end{equation}
satisfying
\begin{equation}\begin{array}{ccl}
\textrm{tr}(P)=1\\
P^2=P.
\end{array}
\end{equation}
This mapping can be generalized for any pure qutrit in $\mathbb{A}^3$, where $\mathbb{A}$ is a composition algebra.  As the resulting density matrix is rank one, satisfying $P^{\#}=0$, one can map any pure qutrit in $\mathbb{A}^3$ to a rank one matrix of $J_3^{\mathbb{A}}$, hence, a point of $\mathbb{AP}^2$.  In the case of the octonions, this is leads to a successful construction of the projective plane $\mathbb{OP}^2$ (Cayley-Moufang plane) \cite{24}, for which the standard construction using only equivalence classes of elements of $\mathbb{A}^3$ fails.  The failure of the standard construction is due to the identification $v\sim \lambda v, v\in \mathbb{A}^3$ requiring associativity to be an equivalence relation \cite{24,53}.  Therefore, for non-associative composition algebras, the rank one operator representation for qutrits is essential in order to recover a sensible theory of quantum mechanics.  Hence, quantum mechanics over non-associative composition algebras is necessarily expressed in the Jordan formulation of quantum mechanics \cite{4,25,40,59}.  In such theories of quantum mechanics a qutrit projective space is maximal, in the sense that there exist topological obstructions preventing the construction of higher projective $n$-spaces $\mathbb{AP}^n, n > 2$, over non-associative composition algebras \cite{53}.\\
\indent As $SU(3)$ and $SL(3,\mathbb{C})$ are the automorphism and reduced structure groups for $J_3^{\mathbb{C}}$, respectively, we will associate the LOCC and SLOCC groups for any qutrit over composition algebra $\mathbb{A}$ with $\textrm{Aut}(J_3^{\mathbb{A}})$ and $\textrm{Str}_0(J_3^{\mathbb{A}})$.  For composition algebras $\mathbb{A}=\mathbb{R},\mathbb{C},\mathbb{H},\mathbb{O},\mathbb{O}_s$, this gives $SO(3)$, $SU(3)$, $USp(6)$, $F_4$ and $F_{4(4)}$ as LOCC groups and $SL(3,\mathbb{R})$, $SL(3,\mathbb{C})$, $SL(3,\mathbb{H})\cong SU^*(6)$, $E_{6(-26)}$, $E_{6(6)}$ as SLOCC groups, respectively.   For $D=5$ supergravities, $\textrm{Aut}(J_3^{\mathbb{A}})$ and $\textrm{Str}_0(J_3^{\mathbb{A}})$ act as rotation and U-duality groups, respectively \cite{4,13,14,20,21}.  This establishes a correspondence between qutrit SLOCC groups over composition algebras and U-duality groups for $D=5$, $\mathcal{N}=8$ and $\mathcal{N}=2$ magic supergravities.

\begin{table} 
\begin{center}
  \begin{tabular}[h!]{ | c | c | c | }
    \hline
    \multicolumn{3}{|c|}{Qudit SLOCC Groups} \\
    \hline
    Composition Algebra $\mathbb{A}$ & Qutrit SLOCC & Qubit SLOCC \\ \hline
    $\mathbb{R}$ & $SL(3,\mathbb{R})$ & $SL(2,\mathbb{R})=\textrm{Spin}(2,1)$ \\ \hline
    $\mathbb{C}$ & $SL(3,\mathbb{C})$ & $SL(2,\mathbb{C})=\textrm{Spin}(3,1)$ \\ \hline
    $\mathbb{H}$ & $SL(3,\mathbb{H})$ & $SL(2,\mathbb{H})=\textrm{Spin}(5,1)$  \\ \hline
    $\mathbb{O}$ & $E_{6(-26)}$ & $SL(2,\mathbb{O})=\textrm{Spin}(9,1)$ \\ \hline 
    $\mathbb{O}_s$ & $E_{6(6)}$ & $SL(2,\mathbb{O}_s)=\textrm{Spin}(5,5)$ \\ \hline
    $\mathbb{O}_{\mathbb{C}}$ & $E_{6}(\mathbb{C})$ & $SL(2,\mathbb{O}_{\mathbb{C}})=\textrm{Spin}(10,\mathbb{C})$ \\ \hline
  \end{tabular}
\end{center}
\caption{SLOCC groups for qutrits and qubits over composition algebras.}
\label{tab:qutrittable}
\end{table}

Single qutrit gates are typically given by unitary transformations \cite{4,34,39,41}.  However, just as for qubits, any invertible linear transformation can serve as a gate for reversible computations \cite{38}.  We can construct qutrit gates by embedding qubit gates and nesting transformations appropriately.  For example, the Hadamard and Pauli gates can be embedded in $3\times 3$ Hermitian gates $M_1, M_2, M_3$ of the form
\begin{eqnarray}
\left(\begin{array}{ccc} 1 & 0 & 0 \\ 0 & r_1 & A_1 \\ 0 & \overline{A}_1 & r_2 \end{array}\right), \left(\begin{array}{ccc} r_3 & 0 & \overline{A}_2 \\ 0 & 1 & 0 \\ A_2 & 0 & r_4 \end{array}\right), \left(\begin{array}{ccc} r_5 & A_3 & 0 \\ \overline{A}_3 & r_6 & 0 \\ 0 & 0 & 1 \end{array}\right),\nonumber
\end{eqnarray} 
where $r_i\in\mathbb{R}$ and $A_i\in\mathbb{A}$ and gate $M_i$ leaves the standard idempotent $P_i$ invariant.  Requiring that $r_i=-r_{i+1}$ and $r_i^2+A\overline{A}=1$, yields unitary gates $M\in\textrm{Aut}(J_2^{\mathbb{A}})\subset\textrm{Aut}(J_3^{\mathbb{A}})$.  SLOCC gates are recovered by requiring $\textrm{det}(M)=1$, giving $M\in\textrm{Str}_0(J_3^{\mathbb{A}})$.  In general, such $M$ gates are in $\textrm{Str}(J_3^{\mathbb{A}})$ and general transformations of elements $J\in J_3^{\mathbb{A}}$ are recovered via nesting
\begin{equation}
J'=M_k(M_j(M_iJM_i^{\dagger})M_j^{\dagger})M_k^{\dagger}.
\end{equation}
Other unitary qutrit transformations include $\textrm{Aut}(J_2^{\mathbb{A}})\subset\textrm{Aut}(J_3^{\mathbb{A}})$ rotations
\begin{eqnarray}
\left(\begin{array}{ccc} e^{\overline{l}\theta} & 0 & 0 \\ 0 & e^{l\theta} & 0 \\ 0 & 0 & 1 \end{array}\right)
\end{eqnarray} 
\begin{eqnarray}
\left(\begin{array}{ccc} \textrm{cos}\theta & l\thinspace\textrm{sin}\theta & 0 \\ \overline{l}\thinspace\textrm{sin}\theta & \textrm{cos}\theta & 0 \\ 0 & 0 & 1 \end{array}\right),
\end{eqnarray}
and $\textrm{Str}_0(J_2^{\mathbb{A}})\subset\textrm{Str}_0(J_3^{\mathbb{A}})$ boosts
\begin{eqnarray}
\left(\begin{array}{ccc} e^{\beta} & 0 & 0 \\ 0 & e^{-\beta} & 0 \\ 0 & 0 & 1 \end{array}\right), \left(\begin{array}{ccc} \textrm{cosh}\beta & \overline{l}\thinspace\textrm{sinh}\beta & 0 \\ l\thinspace\textrm{sinh}\beta & \textrm{cosh}\beta & 0 \\ 0 & 0 & 1 \end{array}\right).
\end{eqnarray}
In \cite{45}, it was shown that for $J_3^{\mathbb{O}}$ one can recover all of $\textrm{Str}_0(J_3^{\mathbb{O}})=E_{6(-26)}$ via suitable nesting of such rotations and boosts.  Since $J_3^{\mathbb{A}}\subseteq J_3^{\mathbb{O}}$ for any composition division algebra $\mathbb{A}$, this applies to other $\textrm{Str}_0(J_3^{\mathbb{A}})$ as well.  It then only remains to show $E_{6(6)}$ can be recovered by nesting of rotations of boosts over the split-octonions, using arguments similar to those given Manogue and Dray in the case of $E_{6(-26)}$ \cite{45}.  Such nesting permits the construction of general SLOCC gates for qutrits over any composition algebra $\mathbb{A}$, using embedded qubit SLOCC gates.  In the corresponding $D=5$, $\mathcal{N}=8$ and $\mathcal{N}=2$ magic supergravities, the nested rotations and boosts yield general U-duality transformations.

\begin{figure}[t]
	\centering
		\includegraphics[width=0.75\textwidth]{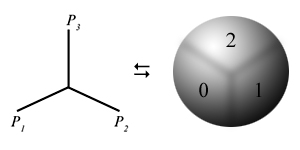}
	\caption{An projective basis for $\mathbb{AP}^2$ as a ternary computational basis.}
	\label{fig:fuzzyqutrit}
\end{figure}

Given a black hole in $D=5$, $\mathcal{N}=8$ and $\mathcal{N}=2$ magic supergravity, we can associate an element $J\in J_3^{\mathbb{A}}$ with spectral decomposition
\begin{equation}
J=\lambda_1P_1+\lambda_2P_2+\lambda_3P_3\quad\quad \lambda_i\in\mathbb{R},
\end{equation} 
where $P_i$ are orthonormal rank one idempotents of $J_3^{\mathbb{A}}$.  Identifying $P_1,P_2,P_3$ with pure qutrit density matrices allows one to interpret 1/4-BPS and 1/8-BPS black hole charge vectors as observables on three orthogonal pure qutrit states.  The eigenvalues are the values the observable takes on three coordinate charts of $\mathbb{AP}^2$.  The automorphism group $\textrm{Aut}(J_3^{\mathbb{A}})\cong\textrm{Isom}(\mathbb{AP}^2)$ preserves pure qutrit normalization while $\textrm{Str}_0(J_3^{\mathbb{A}})$ acts as a SLOCC group preserving the black hole entropy $S=\pi\sqrt{|\textrm{det}(J)|}=\pi\sqrt{|\lambda_1\lambda_2\lambda_3|}$.  This agress with the interpretation of $D=5$ black holes preserving 1/8 supersymmetry as bound states or intersections of basic constituent $p$-branes with charges $Q_1,Q_2,Q_3$, for which the entropy law yields $S=\pi\sqrt{|Q_1Q_2Q_3|}$ \cite{50}.  

As an illustrative example, consider the case of $D=5$, $\mathcal{N}=8$ supergravity, where black hole charge vectors are in $J_3^{\mathbb{O}_s}$, the $\textbf{27}$ of $E_{6(6)}$. Given a qubit SLOCC gate $S\in SL(2,\mathbb{O}_s)=\textrm{Spin}(5,5)$, we can construct a $3\times 3$ qutrit SLOCC gate
\begin{equation}
M=\left(\begin{array}{cc} S & 0 \\ 0 & 1\end{array}\right) \in\textrm{Str}_0(J_3^{\mathbb{O}_s})=E_{6(6)}.
\end{equation}
Acting on a black hole charge vector $J\in J_3^{\mathbb{O}_s}$ imposes the block structure
\begin{equation}
J=\left(\begin{array}{cc} V & \psi \\ \psi^{\dagger} & c\end{array}\right)
\end{equation}
where $V\in J_2^{\mathbb{O}_s}$ transforms as a ten-dimensional vector, $\psi\in \mathbb{O}^2_s$ as a sixteen-dimensional spinor, and $c\in\mathbb{R}$ a scalar.  As $\textrm{Spin}(5,5)$ is the double cover of $SO(5,5)$, choosing one of the three embeddings of a qubit SLOCC gate into a $3\times 3$ qutrit SLOCC gate leads to the decomposition under the T-duality group $SO(5,5)$: $\textbf{27}\rightarrow \textbf{10}+\textbf{16}+\textbf{1}$.  At the group level, this corresponds to the three embeddings of $\textrm{Spin}(5,5)$ in $E_{6(6)}$, giving three such decompositions of the $\mathbf{27}$.  In the projective plane $\mathbb{OP}^2_s$, each copy of $\textrm{Spin}(5,5)$ transforms two orthogonal rank one projectors (hence a projective line), while leaving the third invariant.  This is seen by noting lines in the plane $\mathbb{OP}^2_s$ can be represented by rank two elements \cite{25}
\begin{equation}
P_1+P_2\quad\rightarrow\quad\textrm{line in }\mathbb{OP}^2_s.
\end{equation}
Acting on a diagonalized element of $J_2^{\mathbb{O}_s}$ with $M\in\textrm{Spin}(5,5)\subset E_{6(6)}$ gives
\begin{equation}
MDM^{\dagger}=\lambda_1MP_1M^{\dagger}+\lambda_2MP_2M^{\dagger}+\lambda_3P_3
\end{equation}
leaving a rank one idempotent invariant, while transforming the projective line $l_{12}=P_1+P_2$.  The other two embeddings of the qubit gate $S\in\textrm{Spin}(5,5)$ accordingly transform the remaining two lines $l_{13}=P_1+P_3$ and $l_{23}=P_2+P_3$.
  Quantum computationally, this means an embedded $\textrm{Spin}(5,5)$ acts on a corresponding qubit subspace of the qutrit state space $\mathbb{OP}^2_s$, and by nesting transformations from different $\textrm{Spin}(5,5)$ subgroups a general qutrit transformation is recovered.  Similar results apply to black holes in $D=5$, exceptional $\mathcal{N}=2$ magic supergravity, where the relevant groups are $SO(9,1)$, $\textrm{Spin}(9,1)$ and $E_{6(-26)}$.

\section{Conclusion}
It was shown that 1/2-BPS black string and hole charge vectors in $D=5,6$, $\mathcal{N}=8$ and $\mathcal{N}=2$ magic supergravities can be identified with qubits and qutrits over composition algebras, using the Jordan formulation of quantum mechanics.  This allows one to interpret U-duality groups as SLOCC groups for generalized qubits and qutrits.  Moreover, 1/4 and 1/8-BPS black hole charge vectors were identified with observables on orthogonal qudit states and such an interpretation found agreement with results of multi-charge BPS solutions in toroidally compactified M-theory.  In the case of M-theory on $T^6$, corresponding to $D=5$, $\mathcal{N}=8$ supergravity, the qutrit interpretation provided a novel view of the decomposition of the $\textbf{27}$ of $E_{6(6)}$, under its T-duality subgroup $SO(5,5)$, found to arise from choosing an embedding of qubit SLOCC gates into qutrit SLOCC gates.

Further research into the black hole/qudit correspondence might find applications for BPS black strings and holes in generalized spin chains, as spin-1/2 and spin-1 chains are currently used for both quantum teleporation and transfer of qudit states \cite{60,61}.  Spin chains over non-associative composition algebras, with $SO(5,5)$ and $SO(9,1)$ SLOCC symmetry, might even prove helpful in building novel string bit models \cite{62} for superstrings.  In such a case, superstrings would arise from infinitely long polymers of BPS black string bits.  

Other possible applications of the black hole/qudit correspondence may lie in the study of scattering amplitudes in twistor space \cite{63,64}, where the compact and non-compact complex Hopf fibrations are already in use, mapping spinors to lightlike vectors in (3,1) and (2,2) signature spacetimes. Perhaps the use of more general Hopf fibrations, such as those used for qubits over non-associative composition algebras, will provide deeper insights into the structure of these amplitudes and their interpretation within M-theory.  

Surely, there are many other unforeseen applications, and it appears certain that the black hole/qudit correspondence will continue to illuminate facets of both quantum gravity and quantum information theory and their fascinating interrelationship.

\end{document}